\documentclass[iop,apjl]{emulateapj}
\submitted{accepted for publication in the Astrophysical Journal Letters}
\def\ergs{ergs~s$^{-1}$}
\def\ergcms{ergs$^{-1}$~cm$^{-2}$~s$^{-1}$}

\def\xa{X41.4+60}
\def\xb{X42.3+59}

\begin{document}

\title{Identification of the X-Ray Thermal Dominant State in an Ultraluminous X-Ray Source in M82}

\author{Hua Feng\altaffilmark{1}, \& Philip Kaaret\altaffilmark{2}}

\altaffiltext{1}{Department of Engineering Physics and Center for Astrophysics, Tsinghua University, Beijing 100084, China}
\altaffiltext{2}{Department of Physics and Astronomy, University of Iowa, Van Allen Hall, Iowa City, IA 52242, USA}

\shortauthors{Feng and Kaaret}
\shorttitle{Thermal dominant state in a ULX in M82}

\begin{abstract}
The thermal dominant state in black hole binaries (BHBs) is well understood but rarely seen in ultraluminous X-ray sources (ULXs). Using simultaneous observations of M82 with Chandra and XMM-Newton, we report the first likely identification of the thermal dominant state in a ULX based on the disappearance of X-ray oscillations, low timing noise, and a spectrum dominated by multicolor disk emission with luminosity varying to the 4th power of the disk temperature. This indicates that ULXs are similar to Galactic BHBs. The brightest X-ray spectrum can be fitted with a relativistic disk model with either a highly super-Eddington ($L_{\rm disk}/L_{\rm Edd} = 160$) non-rotating black hole or a close to Eddington ($L_{\rm disk}/L_{\rm Edd} \sim 2$) rapidly rotating black hole.  The latter interpretation is preferred, due to the absence of such highly super-Eddington states in Galactic black holes and active galactic nuclei, and suggests that the ULX in M82 contains a black hole of 200-800 solar masses with nearly maximal spin. On long timescales, the source normally stays at a relatively low flux level with a regular 62-day orbital modulation and occasionally exhibits irregular flaring activity. The thermal dominant states are all found during outbursts. 
\end{abstract}


\section{Introduction}

Ultraluminous X-ray sources (ULXs) are non-nuclear X-ray sources with apparent luminosities above the Eddington limit of stellar-mass black holes (BHs). Variable ULXs are black hole binaries (BHBs) and may harbor intermediate-mass BHs \citep{col99,mak00}. The emission of BHBs has been classified into four states based on spectral and timing properties: the quiescent, hard, thermal dominant (TD), and steep power-law states \citep{mr06,rem06}. The TD state is the best understood and is well described by the standard accretion disk model \citep{sha73}.  The emergent X-ray spectrum is described by a multicolor disk (MCD) model with two parameters: the disk inner radius ($R_{\rm in}$) and the temperature ($T_{\rm in}$) at that radius. In the TD state, $R_{\rm in}$ is constant and the accretion disk is thought to extend all the way to the ``innermost stable circular orbit'' (ISCO) around the BH \citep{tak97}. The ISCO radius depends solely on the mass and spin of the BH. Therefore, a TD spectrum from an accreting BH can be used to shed light on its mass and spin. 

The TD state has been found during outbursts of many BHBs and exhibits specific properties. The disk bolometric luminosity is $L_{\rm disk} = 4 \pi \sigma R_{\rm in}^2 T_{\rm in}^4$. For constant $R_{\rm in}$, a 4th power relation between the luminosity and the disk inner temperature, $L_{\rm disk} \propto T_{\rm in}^4$, is observed \citep{kub04}. Due to up scattering of disk photons in the corona, sometimes the observed exponent is lower than 4, but can be recovered by applying hardening correction \citep{mcc09}. Moreover, sources in the TD state have low levels of short term variability, with very weak or absent narrow band timing noise (quasi-periodic oscillations; QPOs) and weak broad band power continuum \citep{mr06,rem06}.

Spectral surveys of ULXs in nearby galaxies show they are rarely in the TD state \citep{fen05,sto06,win06,sor09}. The spectra of a few ULXs are super-soft and can be modeled by an MCD with an inner temperature close to 0.1 keV. However, repeated observations revealed an evolution pattern inconsistent with the $L_{\rm disk} \propto T_{\rm in}^4$ relation ruling out interpretation as thermal dominant disk emission \citep{liu08}. Some other ULXs show soft excesses in the spectra that could be fitted by cool disk emission \citep{kaa03,mil04}. However, the cool disk is not the dominant component in their spectra, thus, the sources are not in the TD state. Finding a ULX in the TD state would demonstrate that the ULX has emission states similar to Galactic BHBs and allow inference of the BH mass and spin.

The starburst galaxy M82 contains one of the most luminous ULXs, CXOM82 J095550+694047 \citep[=\xa;][]{kaa01,mat01}, in nearby galaxies. The ULX was first identified with Chandra at a luminosity higher than the Eddington limit of a $500 M_{\odot}$ BH \citep{kaa01,mat01}. On the sky, it lies near a super star cluster, which was speculated to be the birth place of a massive BH \citep{por04}. Low frequency QPOs and broadband timing noise, detected in the central region of M82 \citep{str03,dew06,muc06} and later confirmed to originate from this ULX \citep{fen07}, suggest that the ULX harbors a massive BH. \citet{kin05} also suggests that the ULX contains a massive BH which is the nucleus of a satellite galaxy merging with M82.

Positive identification of the emission states requires both timing and spectral information. Here, we describe simultaneous observations exploiting the high angular resolution of Chandra to isolate the ULX spectrum from diffuse emission and nearby sources and the large collecting area of XMM-Newton to obtain timing information.  We use the joint timing and spectral information to identify the ULX emission states and unveil the nature of the source.

\section{Results from Joint Chandra/XMM-Newton Observations}

Three Chandra (ObsIDs 10027, 10025, and 10026) and XMM-Newton (ObsIDs 0560590101, 0560590201, and 0560590301) observations were performed on 2008 October 4, 2009 April 17 and 29, respectively. Each Chandra observation has an effective exposure of about 18 ks and overlapped with an XMM-Newton observation that lasted 30, 40, and 50 ks, respectively.

To reduce the effects of pileup, the Chandra observations placed the ULX $3.57\arcsec$ off the optical axis, so the source covers multiple pixels, and the CCD was operated in a 1/8th sub-array mode to reduce the readout frame time. Using the {\tt pileup\_map} tool in CIAO 4.1.2, we measured that the ULX resulted in 0.25 counts per frame ($\sim$13\% pileup fraction) on the brightest 3-by-3 pixel island in the observation on 2008 Oct 4, and found that about 85\% of the total events were recorded on pixel islands with counts per frame larger than 0.06 ($\sim$3\% pileup fraction). Therefore, pileup must be taken into account in the spectral fitting. We note that the other two observations have similar, mild pileup, which can be corrected by applying a pileup model in the spectral fitting.

Each Chandra spectrum was extracted from a source region found by wavelet detection. Instrument responses were calculated using the calibration files in CALDB 4.1.3. The spectrum channels were re-grouped by a factor of 8 in 0.3-1 keV, 4 in 1-4 keV, 8 in 4-6 keV and 16 in 6-8 keV, respectively and were fitted in XSPEC 12.5 to a power-law model or MCD model (DISKBB) subject to interstellar absorption and CCD pileup with background subtracted from a nearby, source free region. To minimize contamination from diffuse emission and piled events, we fitted the data in the energy range from 0.7 to 7 keV. During the fits, the pileup grade migration parameter $\alpha$ was free, and $f_{\rm psf}$ (the fractional events in the spectrum being treated for pileup) was frozen at 0.85 based on the estimate discussed above and is also the best-fit value found from the fits.

%%%%%%%%%%%%%%%%%%%%%%%%%%%%%%%%%%%%%%%%%%%%%%%%%%%%%%%%%%%%%%%%%%%%%%%%%%
\begin{deluxetable}{lccc}
\tablecolumns{4}
\tablewidth{0pc}
\tablecaption{Spectral parameters of \xa\ fitted with the MCD model subject to interstellar absorption and CCD pileup.
\label{tab:fit}}

\tablehead{
\colhead{Parameters} & \colhead{2008 Oct 04} & \colhead{2009 Apr 17} & \colhead{2009 Apr 29}}

\startdata
$\alpha$ & $0.18 \pm 0.05$ & $0.26 \pm 0.06$ & $>0.29$ \\
$N_{\rm H}$ & $1.13 \pm 0.04$ & $1.07 \pm 0.04$ & $1.05 \pm 0.07$ \\
$T_{\rm in}$ & $1.52 \pm 0.09$ & $1.44 \pm 0.09$ & $1.10_{-0.04}^{+0.21}$ \\
$R_{\rm in}$ & $339 \pm 24$ & $330 \pm 28$ & $311_{-80}^{+24}$ \\
$f_{\rm X}$ & $5.3 \pm 0.8$ & $4.3 \pm 0.5$ & $1.2_{-0.05}^{+0.27}$ \\
$L_{\rm disk}$ & $7.9 \pm 2.0$ & $6.0 \pm 1.8$ & $1.8_{-1.0}^{+1.4}$ \\
$\chi^2/{\rm dof}$ & 67.8/70 & 79.0/68 & 42.5/53
\enddata

\tablecomments{$\alpha$ is the pileup grade migration probability. $N_{\rm H}$ is the absorption column density in $10^{22}$ cm$^{-2}$. $T_{\rm in}$ is the disk inner temperature in keV. $R_{\rm in}$ is the disk inner radius in km calculated assuming a distance of 3.63 Mpc and a disk inclination angle of zero. $f_{\rm X}$ is the absorbed flux in 1-7 keV corrected for pileup in $10^{-11}$~\ergcms. $L_{\rm disk}$ is the bolometric disk luminosity in $10^{40}$~\ergs. All errors are quoted at 90\% confidence level.}
\end{deluxetable}
%%%%%%%%%%%%%%%%%%%%%%%%%%%%%%%%%%%%%%%%%%%%%%%%%%%%%%%%%%%%%%%%%%%%%%%%%%

%%%%%%%%%%%%%%%%%%%%%%%%%%%%%%%%%%%%%%%%%%%%%%%%%%%%%%%%%%%%%%%%%%%%%%%%%%
\begin{figure}
\centering
\includegraphics[width=0.4\textwidth]{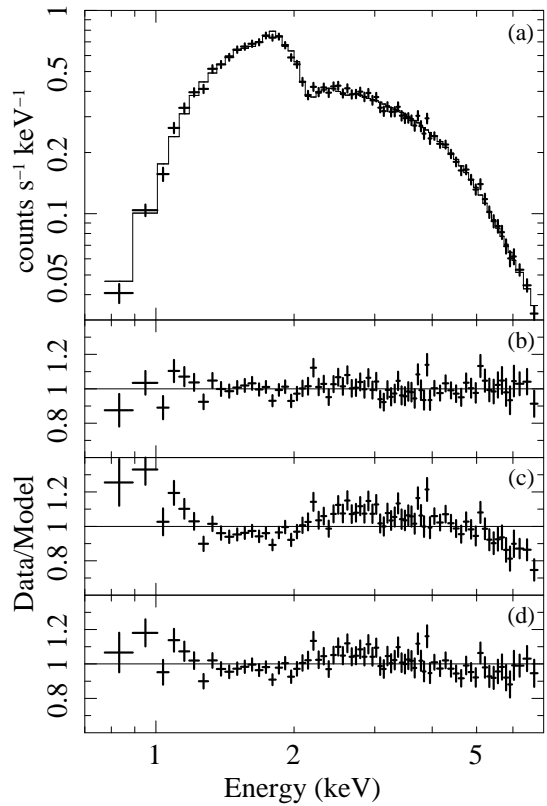}
\caption{The Chandra data spectrum of the ULX on 2008 Oct 04 and comparisons between the MCD and power-law model. (a) The data spectrum and the best-fit MCD model modified by interstellar absorption and CCD pileup (see Table~\ref{tab:fit} for parameters). (b) Data to model ratio for the MCD model ($\chi^2/{\rm dof} = 67.8/70$) (c) Data to model ratio for the power-law model ($\chi^2/{\rm dof} = 201.8/71$) with the pileup parameter $\alpha$ fixed at its best value found from the MCD model. (d) Data to model ratio for the power-law model ($\chi^2/{\rm dof} = 102.1/70$) with $\alpha$ free during the fit, which, however, converged to an unphysical value ($\alpha = 0$) indicative of a zero pileup fraction. Therefore, the MCD model is favored and the power-law model is ruled out from the spectral fitting. 
\label{fig:db_po}}
\end{figure}
%%%%%%%%%%%%%%%%%%%%%%%%%%%%%%%%%%%%%%%%%%%%%%%%%%%%%%%%%%%%%%%%%%%%%%%%%%

We found that the MCD model provides an adequate fit to the data, while the power-law model does not. No reasonable pileup parameters can be found with the power-law model; the $\alpha$ parameter always goes to zero, indicating the pileup fraction in the spectrum is zero, which is contradicted by the pile-up estimates. For the 2008 Oct 04 observation, the fit is not acceptable with $\chi^2 = 102.1$ for 70 degrees of freedom (dof). If we fix $\alpha$ at its best-fit value found from the disk model, the goodness of fit is even worse, with $\chi^2/{\rm dof} = 201.8/71$. In contrast, the MCD model provides a good fit $\chi^2/{\rm dof} = 67.8/70$ and a reasonable value of $\alpha$. Figure~\ref{fig:db_po} shows the spectrum from the Chandra observation on 2008 Oct 04 and the comparison between the MCD and power-law model. Please note that the data/model ratios from the two power-law models are not evenly distributed.

%%%%%%%%%%%%%%%%%%%%%%%%%%%%%%%%%%%%%%%%%%%%%%%%%%%%%%%%%%%%%%%%%%%%%%%%%%
\begin{figure}
\centering
\includegraphics[width=0.4\textwidth]{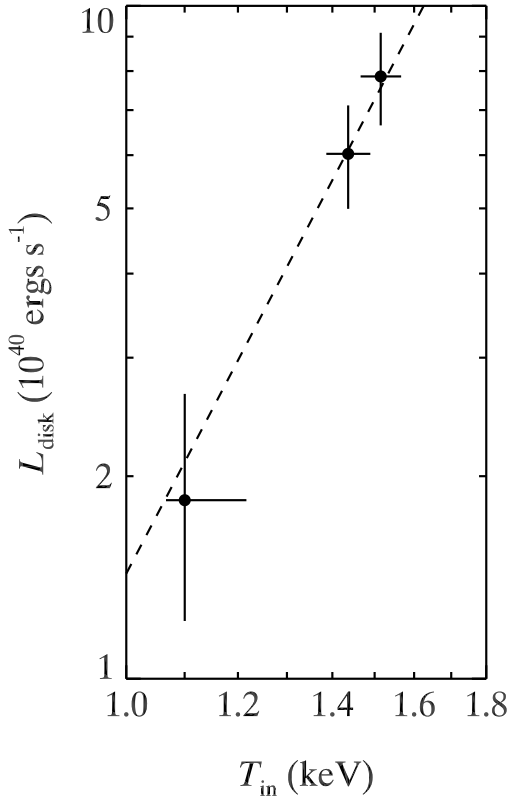}
\caption{Bolometric disk luminosity versus the disk temperature at inner radius of \xa\ derived from the three Chandra observations by fitting the data to the MCD model subject to interstellar absorption and CCD pileup. A constant disk inner radius is derived from the fits, as a consequence of which, the disk luminosity is scaled with the inner temperature to a power of 4. The dashed line indicates a best-fit 4th power-law relation to the data. The errors are at the 1 $\sigma$ level.
\label{fig:lt}}
\end{figure}
%%%%%%%%%%%%%%%%%%%%%%%%%%%%%%%%%%%%%%%%%%%%%%%%%%%%%%%%%%%%%%%%%%%%%%%%%%

The MCD model also provides better fits than power-law for the other two observations, see Table~\ref{tab:fit}. For the observation on 2009 Apr 17, 26.3 of $\chi^2$ arise from instrumental response features on individual channels at 2.1 and 3.2 keV. Ignoring these two channels does not change the spectral parameters at all. We excluded them because we are only interested in the broad continuum. For the observation on 2009 Apr 29, possible emission line features are seen at energies from 4.5 to 7 keV. For the same reason, we only use 0.7-4.5 keV for this observation. We note that for the other two observations the spectral parameters derived from the 0.7-4.5 keV range are completely consistent with from the 0.7-7 keV range. Although adding a power-law component to the model does not improve the fits, we did that to place an upper limit on the power-law flux with the power-law photon index bounded in the range 2.1-4.8 as found in the TD state. The upper limits on a power-law component in 2-20 keV are estimated to be 4\%, 17\%, and 20\%, respectively. This is consistent with the definition of the TD state.

Power spectra were calculated from XMM-Newton data using combined PN and MOS events in a circular region around the source with a radius of $18\arcsec$ from the common good time intervals without timing gaps \citep[following][]{fen07}. We searched for QPOs from 1 to 1000 mHz in various energy ranges, but did not find significant signals. We repeated the search in the half of the circular source region where contamination from sources other than the ULX is minimized \citep[region A in Fig.\ 2 of][]{fen07}, but still found nothing. The power spectra are consistent with that from the white noise. The 3$\sigma$ upper limits of the total noise power (in units of rms/mean) in 1-1000 mHz in the energy range of 0.3-10 keV are 6.3\%, 6.6\%, and 6.9\%, respectively from the three observations. 

These XMM-Newton observations are able to detect QPOs at the strengths previously observed. The disappearance of QPOs and the low noise level suggest that the source was not in the hard state. The Chandra energy spectra show that they are incompatible with a power-law form, but can be adequately modeled with a dominated MCD model with a constant inner radius. Figure~\ref{fig:lt} shows that the bolometric luminosity from the accretion disk varies with the 4th power of the temperature. The correlation coefficient between $\log L_{\rm disk}$ and $\log T_{\rm in}$ is 0.9995 with a chance probability of 0.02 from the three points. All of the emission properties, low timing noise, spectrum dominated by an MCD with low power-law fraction, and luminosity proportional to disk temperature to the 4th power are consistent with the source being in the TD state and are strong evidence in favor of the interpretation of the energy spectrum as emission from an optically thick accretion disk.

%%%%%%%%%%%%%%%%%%%%%%%%%%%%%%%%%%%%%%%%%%%%%%%%%%%%%%%%%%%%%%%%%%%%%%%%%%
\begin{figure}
\centering
\includegraphics[width=0.5\textwidth]{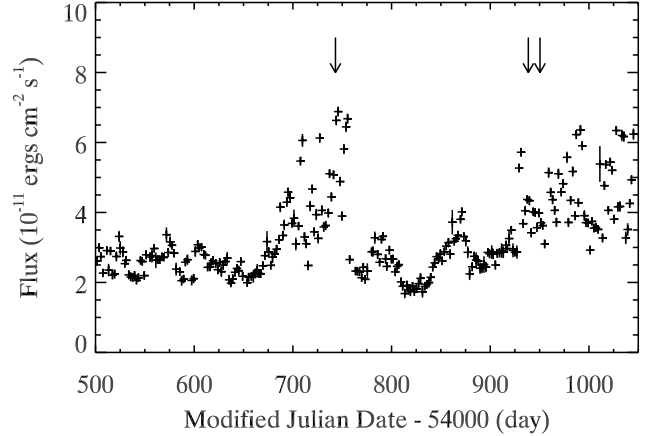}
\caption{Lightcurve of the whole M82 galaxy observed with RXTE from 2008 Feb 5 to 2009 Aug 2. The crosses are the observed X-ray flux in the energy range of 2-10 keV, which is dominated by \xa. The source spends most of its time at a relatively low flux level, around which the flux is modulated at a period of 62 days, and occasionally shows flaring activity lasting days to months. The three arrows indicate the times of the joint Chandra and XMM-Newton observations, in which QPOs disappeared and the source was identified as in the TD state. 
\label{fig:lc}}
\end{figure}
%%%%%%%%%%%%%%%%%%%%%%%%%%%%%%%%%%%%%%%%%%%%%%%%%%%%%%%%%%%%%%%%%%%%%%%%%%

The lightcurve from RXTE for the whole M82 galaxy is shown in Figure~\ref{fig:lc} from 2008 February 5 to 2009 August 2. The Chandra/XMM-Newton described here all occurred when the flux from M82 was well above its nominal level.  The ULX was, by far, the brightest source in the three Chandra observations. Therefore, these outbursts from M82 are dominated by the ULX, with a minor contribution from another source (\xb) active only in the first joint observations. Thus, the TD states were all observed during bright outbursts of the ULX.

Only two other Chandra observations have manageable pileup.  An observation on 2005 February 4 revealed a relatively low X-ray flux and a hard power-law spectrum inconsistent with an MCD model \citep{kaa06} suggesting the hard state.  We previously analyzed an observation from 2007 July 2, but fixed $f_{\rm psf} = 0.38$ based on the assumption that pileup was important only in the 3-by-3 pixel cluster with the highest count rate \citep{kaa09}. Utilizing the new {\tt pileup\_map} tool, we realized that $f_{\rm psf}$ was underestimated. Using the improved procedures described above, we find $\chi^2_{\nu} = 2.2$ for a power-law model and $\chi^2_{\nu} = 1.4$ for a MCD model with parameters similar to those from 2008 Oct 04. Lack of timing information precludes firm conclusion, but the TD state is suggested.

\subsection{Fitting with a Relativistic Disk Model}

The standard MCD model does not include relativistic effects. Such effects are particularly important for rapidly spinning BHs. For accurate estimation of the BH properties, we fitted the spectrum from when the source was brightest, the 2008 observation, with a fully relativistic MCD model, KERRBB \citep{li05}. We set zero torque \citep{li05,mcc06} at the inner edge of the disk, the distance to the source as 3.63 Mpc \citep{fre94}, the hardening factor as 1.7 \citep{mak00}, and switched on self-irradiation and limb-darkening. The pileup parameters were fixed at their best values found from the MCD model. This left five free parameters in the model: the disk inclination, BH mass, spin, accretion rate, and the interstellar absorption column density.

%%%%%%%%%%%%%%%%%%%%%%%%%%%%%%%%%%%%%%%%%%%%%%%%%%%%%%%%%%%%%%%%%%%%%%%%%%
\begin{figure}
\centering
\includegraphics[width=0.4\textwidth]{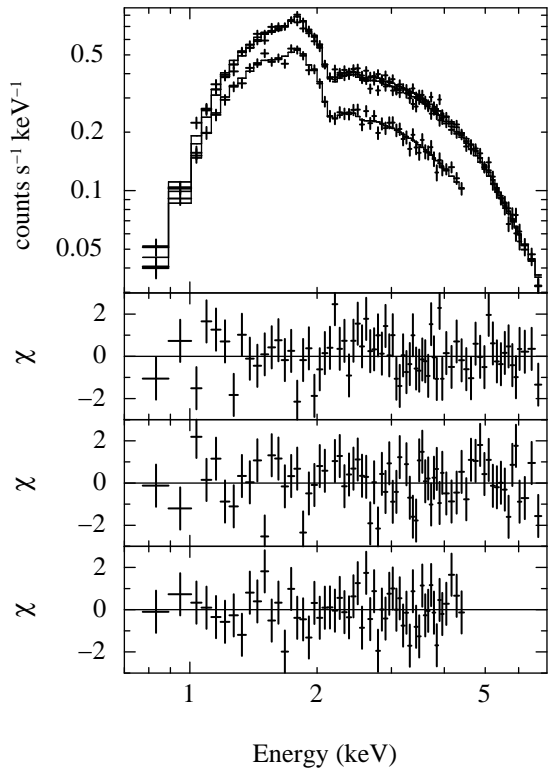}
\caption{Energy spectra of \xa\ from the three Chandra observations and the simultaneous fit to the KERRBB model subject to interstellar absorption and CCD pileup. The spin parameter is fixed at its maximum $a_* = 0.9986$. For other parameters please refer to the discussion of this model in the text. The residues are for the observations on 2008 Oct 4, 2009 Apr 17, and 2009 Apr 29, respectively, from top to bottom and in units of $\sigma$.
\label{fig:kerrbb}}
\end{figure}
%%%%%%%%%%%%%%%%%%%%%%%%%%%%%%%%%%%%%%%%%%%%%%%%%%%%%%%%%%%%%%%%%%%%%%%%%%

To test if the BH is maximally spinning, we fixed the spin parameter at two extreme values $a_* = 0$ and 0.9986, respectively, representing non-spinning and maximally spinning \citep{li05}. For a non-rotating BH, the accretion rate exceeds the Eddington limit by a factor of 160. For Galactic BHBs, the observed Eddington ratio is as high as 7 in V4641 Sgr despite the uncertainty on the distance \citep{mr06}. When the luminosity is high and the radiation pressure is dominant, the thin accretion disk is predicted to be able to exceed the Eddington limit by a factor of a few \citep{beg02}. Therefore, we adopt an upper limit of 10 for the Eddington ratio as being physically reasonable. We note that active galactic nuclei do not appear to exceed the Eddington limit by more than a factor of 10 in any accretion state \citep{war04}. The Eddington ratio for a non-rotating BH is too large and this scenario should be ruled out. In contrast, if the BH is maximally spinning, the derived Eddington ratio is only 2.

Freezing the spin at the maximum allowed value resulted in a best-fit BH mass of 660 solar masses and a 90\% error range of 300-1250 solar masses, a disk inclination angle of at least $60\arcdeg$, and $\chi^2/{\rm dof} = 67.7/70$. The other two observations produce similar results.  We also applied the model to the three observations simultaneously, imposing the same mass, spin, and inclination, and allowing the mass accretion rate and absorption column density to vary individually. Fixing the spin parameter $a_* = 0.9986$ led to a best-fit BH mass of 500 solar masses, a 90\% error range of 300-810 solar masses, an inclination of 59-79 degrees, and $\chi^2/{\rm dof} = 189.0/195$; see Figure~\ref{fig:kerrbb}. The Eddington accretion rate at the highest luminosity observed is 2.5. The inferred Eddington ratio exceeds 10 for $a_* < 0.93$, which we took as the lower limit of the spin. Fixing $a_* = 0.93$, the best-fit BH mass is 430 solar masses with a 90\% error range of 190-570 solar masses, and the inclination is larger than $60\arcdeg$. Thus, the spectral fitting suggests that the ULX contains a rapidly spinning ($a_* > 0.93$) intermediate mass BH of 200-800 solar masses.

\section{Discussion}

Joint Chandra and XMM-Newton observations have enabled us to precisely measure the spectral and timing behavior of this ULX in M82. Disappearance of QPOs and low timing noise, an X-ray spectrum best-fitted with an MCD model, and an $L_{\rm disk} \propto T_{\rm in}^4$ pattern in the spectral evolution provide strong motivation to interpret this behavior in terms of the TD state seen in stellar-mass BHBs. The MCD model indicates that the source spectrum shows spectral curvature at a few keV and thus could also be fitted by more complicated models like a cool optically thick corona which has been proposed recently as an indicator of a new `ultraluminous state' \citep{gla09}. However, a transition from the hard spectrum (incompatible with the steep power-law state) seen at lower luminosities to the ultraluminous state would be difficult to explain and we, therefore, do not discuss this proposed state in detail.

Simply scaling the $L_{\rm disk} \propto T_{\rm in}^4$ pattern to those from Galactic BHBs suggests that the source contains a more massive BH, because the compact object mass is proportional to the square root of the disk luminosity if two BHs have the same disk inner temperature and spin \citep{mak00}. Fitting with a fully relativistic MCD model leads to a consistent result. A conservative estimate of the BH mass by allowing an Eddington ratio of as high as 10 suggests a BH of at least 200 solar masses but less than 800 solar masses. This is coincident with the theoretical estimate of the mass of this BH if it is created in a nearby star cluster by runaway collisions \citep{por04}. The fast spin of the source makes it efficient in extracting gravitational energy.

The spectral fitting also indicates a relatively high inclination angle, 59-79 degrees, of the accretion disk. For a maximally spinning BH of a few hundred solar masses, the relativistic effect has largely eliminated the limb-darkening effect; viewing the disk at a high inclination angle receives almost the same flux as at a low angle. Therefore, the observed high flux is not a problem for a nearly edge-on disk. The X-ray flux from M82 is modulated at a period of 62 days, interpreted as due to orbital motion of this ULX binary, which must contain a giant or super-giant star inferred from the period assuming Roche-lobe overflow \citep{kaa06,kaa07}. With knowledge of the BH mass (assuming 200-800 solar masses) and the 62-day binary orbital period, the binary separation of the ULX can be calculated with the assumption of Roche lobe overflow as $(3-4) \times 10^{13}$~cm, which is insensitive to the mass of the companion star. This is larger than all separations of low-mass BHBs with a dynamical measurement of the mass, and larger than that of GRS 1915+105, which has the largest separation known so far, by a factor of a few. This ULX and GRS 1915+105 share a similarity that they have both been active for many years without quenching to the quiescent state. The large separation and consequently a huge reservoir of accretion mass could be the factor that determines their long-term activity \citep{dee09}. 

\acknowledgments We thank the referee for useful comments and the mission planning teams of Chandra and XMM-Newton for making these observations possible. HF acknowledges funding support from the National Natural Science Foundation of China under grant No.\ 10903004 and 10978001, the 973 Program of China under grant 2009CB824800, and the Foundation for the Author of National Excellent Doctoral Dissertation of China under grant 200935. PK acknowledges support from Chandra grant GO9-0034X.


\begin{thebibliography}{}

\bibitem[Begelman(2002)]{beg02}
Begelman, M.\ C.\ 2002, \apj, 568, L97

\bibitem[Colbert \& Mushotzky(1999)]{col99}
Colbert, E.\ J.\ M., \& Mushotzky, R.\ F.\ 1999, \apj, 519, 89

\bibitem[Deegan et al.(2009)]{dee09}
Deegan, P., Combet, C., \& Wynn, G. A.\ 2009, \mnras, in press (arXiv:0908.2566)

\bibitem[Dewangan et al.(2006)]{dew06}
Dewangan, G.\ C., Titarchuk, L., \& Griffiths, R.\ E.\ 2006, \apj, 637, L21

\bibitem[Feng \& Kaaret(2005)]{fen05}
Feng, H., \& Kaaret, P.\ 2005, \apj, 633, 1052

\bibitem[Feng \& Kaaret(2007)]{fen07}
Feng, H., \& Kaaret, P.\ 2007, \apj, 668, 941

\bibitem[Freedman et al.(1994)]{fre94}
Freedman, W.\ L., et al.\ 1994, \apj, 427, 628

\bibitem[Gladstone et al.(2009)]{gla09}
Gladstone, J.\ C., Roberts, T.\ P., \& Done, C.\ 2009, \mnras, 397, 1836

\bibitem[Kaaret et al.(2003)]{kaa03} 
Kaaret, P., Corbel, S., Prestwich, A.\ H., \& Zezas, A.\ 2003, Science, 299, 365

\bibitem[Kaaret \& Feng(2007)]{kaa07}
Kaaret, P., \& Feng, H.\ 2007, \apj, 669, 106

\bibitem[Kaaret et al.(2009)]{kaa09} 
Kaaret, P., Feng, H., \& Gorski, M.\ 2009, 692, 653

\bibitem[Kaaret et al.(2001)]{kaa01} 
Kaaret, P., Prestwich, A.\ H., Zezas, A., Murray, S.\ S., Kim, D.-W., Kilgard, R.\ E., Schlegel, E.\ M., \& Ward, M.\ J.\ 2001, \mnras, 321, L29

\bibitem[Kaaret et al.(2006)]{kaa06}
Kaaret, P., Simet, M.\ G., \& Lang, C.\ C.\ 2006, \apj, 646, 174

\bibitem[King \& Dehnen(2005)]{kin05}
King, A.\ R., \& Dehnen, W.\ 2005, \mnras, 357, 275

\bibitem[Kubota \& Makishima(2004)]{kub04}
Kubota, A., \& Makishima, K.\ 2004, \apj, 601, 428

\bibitem[Li et al.(2005)]{li05}
Li, L.-X., Zimmerman, E.\ R., Narayan, R., \& McClintock, J.\ E. 2005, \apjs, 157, 335

\bibitem[Liu(2008)]{liu08}
Liu, J.-F.\ 2008, \apjs, 177, 181

\bibitem[Makishima et al.(2000)]{mak00}
Makishima, K., et al.\ 2000, \apj, 535, 632

\bibitem[Matsumoto et al.(2001)]{mat01}
Matsumoto, H., Tsuru, T.\ G., Koyama, K., Awaki, H., Canizares, C.\ R., Kawai, N., Matsushita, S., \& Kawabe, R.\ 2001, \apj, 547, L25

\bibitem[McClintock \& Remillard(2006)]{mr06}
McClintock, J.\ E., \& Remillard, R.\ A.\ 2006, in Compact Stellar X-ray Sources,
ed.\ W.\ H.\ G.\ Lewin \& M.\ van der Klis (Cambridge: Cambridge Univ.\ Press), 157

\bibitem[McClintock et al.(2009)]{mcc09}
McClintock, J.\ E., Remillard, R.\ A., Rupen, M.\ P., Torres, M.\ A.\ P., Steeghs, D., Levine, A.\ M., \& Orosz, J.\ A.\ 2009, \apj, 698, 1398

\bibitem[McClintock et al.(2006)]{mcc06}
McClintock, J.\ E., Shafee, R., Narayan, R., Remillard, R.\ A., Davis, S.\ W., \& Li, L.-X.\ 2006, \apj, 652, 518

\bibitem[Miller et al.(2004)]{mil04}
Miller, J.\ M., Fabian, A.\ C., \& Miller, M.\ C.\ 2004a, \apj, 614, L117

\bibitem[Mucciarelli et al.(2006)]{muc06}
Mucciarelli, P., Casella, P., Belloni, T., Zampieri, L., \& Ranalli, P.\ 2006, \mnras, 365, 1123

\bibitem[Portegies Zwart et al.(2004)]{por04}
Portegies Zwart, S.\ F., Baumgardt, H., Hut, P., Makino, J., \& McMillan, S.\ L.\ W.\ 2004, \nat, 428, 724

\bibitem[Remillard \& McClintock(2006)]{rem06}
Remillard, R.\ A., \& McClintock, J.\ E.\ 2006, \araa, 44, 49 

\bibitem[Shakura \& Sunyaev(1973)]{sha73}
Shakura, N.\ I., \& Sunyaev, R.\ A.\ 1973, \aap, 24, 337

\bibitem[Soria et al.(2009)]{sor09}
Soria, R., Risaliti, G., Elvis, M., Fabbiano, G., Bianchi, S., \& Kuncic, Z.\ 2009, \apj, 695, 1614

\bibitem[Stobbart et al.(2006)]{sto06}
Stobbart, A.-M., Roberts, T.\ P., \& Wilms, J.\ 2006, \mnras, 368, 397

\bibitem[Strohmayer \& Mushotzky(2003)]{str03}
Strohmayer, T.\ E., \& Mushotzky, R.\ F.\ 2003, \apj, 586, L61

\bibitem[Takana \& Lewin(1997)]{tak97}
Takana, Y., \& Lewin, W.\ H.\ G.\ 1997, in X-ray Binaries, ed.\ W.\ H.\ G.\ Lewin, J.\ van Paradijs, \& E.\ P.\ J.\ van den Heuvel (Cambridge Univ.\ Press), 126

\bibitem[Warner et al.(2004)]{war04}
Warner, C., Hamann, F., \& Dietrich, M.\ 2004, \apj, 608, 136

\bibitem[Winter et al.(2006)]{win06}
Winter, L.\ M., Mushotzky, R.\ F., \& Reynolds, C.\ S.\ 2006, \apj, 649, 730

\end{thebibliography}
\end{document}